\documentclass[reviewcopy]{elsarticle}

\usepackage[reviewcopy]{adndt}
\usepackage{longtable}

\usepackage{amsmath}

\usepackage{amssymb}

\biboptions{square,sort&compress}
\bibpunct[]{[}{]}{,}{n}{}{;}
\newcommand{\journaltitle}{Atomic Data and Nuclear Data Tables}
\newcommand{\journalhome}{\texttt{http://www.elsevier.com/locate/jnlabr/yadndt}}
\newcommand{\journalmail}{\texttt{adndt@elsevier.com}}
\newcommand{\cmd}[1]{\texttt{$\backslash${#1}}}
\newcommand{\templatename}{\texttt{tmpadndt.tex}}
\newcommand{\adndtstyle}{\texttt{adndt}}
\newcommand{\adndtbst}{\texttt{adndt.bst}}
\newcommand{\adndtguide}{\texttt{../doc/ADNDdoc.pdf}}
\begin{document}

\begin{frontmatter}

\journal{Atomic Data and Nuclear Data Tables}

%% Author, fill in article title here

\title{Radiative rates for E1, E2, M1, and M2 transitions in  F-like ions with 55 $\le$ Z $\le$ 73}%Template for \normalfont\textsc{Atomic Data and Nuclear Data Tables}}

%% Fill in author list here
  \author[One]{Kanti M. Aggarwal\fnref{}\corref{cor1}}
  \ead{K.Aggarwal@qub.ac.uk}

%  \author[One]{Francis P. Keenan}%\fnref{}}

 % \author[Two]{C. Author} 

  \cortext[cor1]{Corresponding author.}
%  \fntext[X]{{\em e-mail address}: K.Aggarwal@qub.ac.uk}
%  \fntext[Y]{{\em e-mail address}: F.Keenan@qub.ac.uk}

  \address[One]{Astrophysics Research Centre, School of Mathematics and Physics, Queen's University Belfast,\\Belfast BT7 1NN,
Northern Ireland, UK}

%  \address[Two]{Second Address First Line\\
%    Second Address Second Line}

\date{16.12.2002} %please do not use \today, use actual date of version

\begin{abstract}  
Energy levels, radiative rates and lifetimes are reported for 19 F-like ions with 55 $\le$ Z $\le$ 73, among  113 levels of the 2s$^2$2p$^5$, 2s2p$^6$, 2s$^2$2p$^4$3$\ell$, 2s2p$^5$3$\ell$, and 2p$^6$3$\ell$ configurations.   The general-purpose relativistic atomic structure package ({\sc grasp}) has been adopted for the calculations,   and radiative rates (and other associated parameters, such as  oscillator strengths and line strengths) are listed  for all E1, E2, M1, and M2 transitions of the ions. Comparisons are made with  earlier available theoretical and experimental energies, especially for Ba~XLVIII. Nevertheless,  calculations have also been performed with the flexible atomic code ({\sc fac}), and with a much larger configuration interaction with  up to 38~089 levels, for further accuracy assessments, particularly for energy levels. \\ \\

{\em Received}: 9 November 2017, {\em Accepted}: 14 January 2018

\vspace{0.5 cm}
{\bf Keywords:} F-like ions, energy levels, radiative rates, oscillator strengths,  lifetimes
\end{abstract}

\end{frontmatter}

%%% Keywords and subject classification are not used in ADNDT 
%%%\begin{keywords}
%%%Insert list of keywords here.
%%%\end{keywords}

%%% The table of contents should start a new page. This command will
%%% automatically pull all the unstarred \section, \subsection and
%%% \subsubsection titles into the Contents. Starred versions need to be
%%% done manually using
%%%            \addcontentsline{toc}{[[sub]sub]section}{Section title}
%%% at the correct place. Examples are given below.

%%% The lists of data figures and data tables are created automatically
%%% by the \listofDfigures and \listofDtables commands.

\newpage

\tableofcontents
\listofDtables
\listofDfigures
\vskip5pc

%%%% Authors begin text of article here %%%

\section{Introduction}

F-like ions have been of interest for the modelling of fusion plasmas for a long time \cite{plh}, and with the developing ITER project their importance has further increased. The first extensive study for these ions was done by Sampson et al. \cite{samp}, who performed calculations for a wide range of F-like ions with 22 $\le$ Z $\le$ 92 by using their Dirac-Fock-Slater (DFS) code. However, for brevity they reported only limited results (and for only a few ions) for energy levels, oscillator strengths (f-values) and collision strengths ($\Omega$). Unfortunately, most of their data have now been lost. Nevertheless, a few workers after them have performed calculations for a variety of atomic parameters, for a section of these ions. For example, in our earlier study \cite{ak1}, we reported energy levels, radiative rates (A-values), oscillator strengths (f-values), line strengths (S-values), and lifetimes ($\tau$) for 17 F-like ions with 37 $\le$ Z $\le$ 53. Similar data for Z = 36 (Kr~XXVIII \cite{kr}), 54 (Xe~XLVI \cite{xe}) and 74 (W~LXVI \cite{w66a,w66b,w66c}) have also been reported. In this paper we list our results for further 19 ions with 55 $\le$ Z $\le$ 73.

The prior results for these ions, theoretical or experimental, are (mostly) limited to Ba~XLVIII alone. Hutton et al. \cite{rh} have measured spectra for F-like, O-like and N-like barium ions through the electron beam ion trap (EBIT) machine, and their energy levels have been recommended by the NIST (National Institute of Standards and Technology) team \cite{team}, and are available at  the  website {\tt http://www.nist.gov/pml/data/asd.cfm}. Similarly, in a laser produced plasma Feldman et al. \cite{uf2} measured wavelengths for two lines among the lowest three levels of a few F-like ions, and their extrapolated results for Ba~XLVIII are also included in the NIST database. Theoretically, as stated earlier Sampson et al. \cite{samp} performed calculations for a wide range of F-like ions, but did not specifically report energy levels, and data for other parameters was also limited to a few levels/transitions. Using a combined configuration interaction (CI) and many-body perturbation theory (MBPT) approach, Gu \cite{gu} calculated energies for the lowest three levels of F-like ions with Z $\le$ 60. For the same three levels, later on J\"{o}nsson et al. \cite{jag} reported energies and A-values for a wider range of ions with 14 $\le$ Z $\le$ 74, for which they   adopted the general-purpose relativistic atomic structure package ({\sc grasp}) code \cite{grasp2k}, and  included very large CI for the calculations. Using the original version of the same code (GRASP0 \cite{grasp0}), but extensively modified by (one of the authors) P.~H.~Norrington, Khatri et al. \cite{khat} calculated energies for 431 levels of Ba~XLVIII, which belong to the 29 configurations, namely  2s$^2$2p$^5$, 2s2p$^6$, 2s$^2$2p$^4$3$\ell$,  2s2p$^5$3$\ell$, 2p$^6$3$\ell$,   2s$^2$2p$^4$4$\ell$,   2s2p$^5$4$\ell$, 2s$^2$2p$^4$5$\ell$,  and 2s2p$^5$5$\ell$. However, their corresponding results for A-values are limited for transitions from the lowest three levels alone, whereas in a plasma modelling calculation a complete set of data, for all transitions, is preferred. Therefore, their is scope for the extension of their work. 

With the easy and free availability of the atomic structure codes and comparatively cheaper access to computational resources, it has become much easier to generate atomic data for various parameters. Unfortunately, producing reliable data with (some measure of)  accuracy is still not straightforward because many checks are required before having confidence in the reported data. For energy levels, one may assess the accuracy by: (i) comparing results with two different independent codes, (ii) confirming convergence of results by increasing CI, (iii) comparing results with earlier available theoretical data, and (iv) making comparisons with measurements. In spite of such checks and balances, large discrepancies are often observed for (almost) all atomic parameters, as recently highlighted and explained in our paper \cite{atom}. For example,  Goyal  et al. {\cite{jam}} reported energies for  113 levels of the 2s$^2$2p$^5$, 2s2p$^6$,  2s$^2$2p$^4$3$\ell$, 2s2p$^5$3$\ell$, and 2p$^6$3$\ell$ configurations of W~LXVI, for which they adopted two different codes, namely GRASP and FAC, the flexible atomic code of Gu \cite{fac}. However,  for the highest 20 levels the two sets of energies differed by up to $\sim$60~Ryd. Through our independent calculations \cite{w66c}  it turned out that their results with FAC are incorrect. Similarly, for some Ne-like ions differences in energies, between the GRASP and FAC results,  for some levels are up to $\sim$2~Ryd  -- see tables 2--4 of \cite{sam}. Since such a large difference between any two codes is generally neither noted nor expected, one of the two calculations is actually not correct, as explained in our paper \cite{cjp1}. Therefore, comparison of  results with two different  codes is only beneficial when both calculations have been performed with great care.

In most atomic structure calculations, inclusion of CI helps to improve the energy levels, and this is particularly true for ions with comparatively lower atomic numbers. However, there is always a limit up to which the calculations can be performed, and in many cases inclusion of CI beyond a (certain) level is of no real advantage. For F-like ions an extensive inclusion of CI is not very beneficial, as noted in our earlier work \cite{ak1}, although   J\"{o}nsson et al. \cite{jag} have performed very large calculations by including up to 73~000 and 15~000 CSFs (configuration state functions) for the 2s$^2$2p$^5$~$^2$P$^o_{1/2}$ and 2s2p$^6$~$^2$S$_{1/2}$ levels, as a result of which their energies  closely match with the measurements. However, inclusion of such a large CI is not practical for a large number of levels (and ions), particularly when measurements are almost non existent.

Comparison of results with existing data are always useful and considerably help in improving the calculations -- see for example the recent work of Guo et al.  \cite{guo} on W~XL and the references within the paper. However, the problem arises when no prior data exist as noted for some of the Cr-like ions \cite{crlike1}. Similarly, if measurements are available, even for a few levels, it considerably helps to improve the calculations, as was the case for some other Cr-like ions -- see Table~A of \cite{crlike2}.  Unfortunately, for the current F-like ions of interest, theoretical data are (mostly) restricted to the lowest three levels, and the measurements for a few levels are available for only Ba~XLVIII. Therefore, we will make most of the comparisons for this ion alone to make some assessment of the accuracy of our calculated data.

\section{Energy levels}

As in our earlier work \cite{ak1}, we adopt the GRASP0 version which is hosted at the website  {\tt http://amdpp.phys.strath.ac.uk/UK\_APAP/codes.html}.   For the optimisation of the orbitals we  use the option of  `extended average level' (EAL),  in which a weighted (proportional to 2$j$+1) trace of the Hamiltonian matrix is minimised. Additionally, the contributions of Breit and quantum electrodynamic effects (QED) are included, which are important for the heavy ions considered in this work.  Calculations for energy levels and subsequent other parameters have been performed among 501 levels of 38 configurations, namely  2s$^2$2p$^5$, 2s2p$^6$, 2s$^2$2p$^4$3$\ell$, 2s2p$^5$3$\ell$,  2p$^6$3$\ell$, 2s$^2$2p$^4$4$\ell$, 2s2p$^5$4$\ell$, 2p$^6$4$\ell$, 2s$^2$2p$^4$5$\ell$, 2s2p$^5$5$\ell$, and 2p$^6$5$\ell$. However, as in the past, our focus is on 113 levels of the 2s$^2$2p$^5$, 2s2p$^6$, 2s$^2$2p$^4$3$\ell$, 2s2p$^5$3$\ell$, and 2p$^6$3$\ell$ (11) configurations. Furthermore, A-values have been calculated for  four types of transitions, namely electric dipole (E1),  magnetic dipole (M1), electric quadrupole (E2), and magnetic quadrupole (M2). These results are required for the further calculation of lifetimes as well as in plasma modelling.

Our calculated energies with GRASP for all 113 levels for F-like ions with 55 $\le$ Z $\le$ 73 are listed in Tables~1--19.  For the ground level  (2s$^2$2p$^5$~$^2$P$^o_{3/2}$) absolute energies are listed whereas for others  are differences w.r.t. to the ground. Since there is a paucity of prior theoretical and experimental data for these ions, we have performed analogous calculations with FAC ({\tt https://www-amdis.iaea.org/FAC/}), which is also a fully relativistic code and generally yields energy levels of comparable accuracy, but with much more efficiency. With this code three calculations have been performed which are: (i) FAC1, which includes 501 levels of the the same configurations as included with GRASP, (ii) FAC2, which includes 38~089 levels arising from all possible combinations of the (2*5) 3*2, 4*2, 5*2, 3*1 4*1, 3*1 5*1, and 4*1 5*1 configurations, plus those of FAC1, and (iii) FAC3, which includes a total of 72~259 levels, the additional ones arising from the (2*6) 6*1, 7*1, 8*1 and (2*5 3*1) 6*1,  7*1, and  8*1 configurations. Since our FAC1 energies are almost comparable with those with GRASP, in both magnitude and orderings, we list these only for Ba~XLVIII in Table~2. This result was expected because both calculations include the same CI. However, our FAC2 calculations include much more extensive CI and therefore are expected to be more accurate, we include these results for all ions. As was the case earlier for other F-like ions \cite{ak1}, the FAC3 calculations offer no real advantage, because there are no appreciable differences with the FAC2 energies, for the lowest 113 levels under consideration. For this reason we  do not discuss the FAC3 energies any further. Similarly, the results of Khatri et al. \cite{khat} are not included in Table~2 because there are no differences with our present calculations. This is mainly because both calculations adopt the same code and the addition of further 70 levels of the  2p$^6$4$\ell$ and 2p$^6$5$\ell$ (9) configurations in our calculations has an insignificant effect on the 113 levels considered in this work.

As already stated, the contributions of Breit and QED effects are very important and significant for the determination of energy levels for these heavy ions. However, the maximum effect is on the ground levels and therefore, in Table~A along with the Coulomb energies we list contributions of Breit and QED corrections for all ions. Although the combined effect of these corrections on total energies is less than 0.2\%, in absolute terms these are very significant. For example, for Cs~XLVII their respective contributions amount to 9.1+7.6 = 16.7~Ryd, but increase by a factor of 2.5 to 22.7+20.0 = 42.7~Ryd for Ta~LXV. In comparison, the maximum combined contribution to other levels is $\sim$4~Ryd, i.e. less than 10\%.  Similarly, differences between the absolute energies obtained for the ground level with GRASP and FAC increase with increasing Z, as it is only $\sim$1~Ryd for Cs~XLVII but $\sim$6~Ryd for Ta~LXV, and energies with the former are always the {\em lowest}. However, comparative differences between energies from both calculations remain within $\sim$~0.5~Ryd, for all excited levels and ions, as seen in Tables~1--19.

Before we discuss our results in detail, we would like to emphasize on two points. Firstly, the level designations provided in Tables~1--19 may not be unambiguous. This is because some of the levels are highly mixed, and for a few mixing coefficient from a particular level/configuration may dominate in more than one levels. This is a general atomic structure problem, irrespective of the code adopted. Although this has been discussed several times in some of our earlier papers (as well as by other workers), in Table~B we provide the mixing coefficients for the levels of Cs~XLVII, to give some idea of the problem. Among the problematic levels are 12/35, 15/44 and 18/47. We have tried to provide a unique designation for each level, but that is subject to change with other calculations and/or workers. Therefore,  only the $J^{\pi}$ values should be considered definite. Secondly, unlike the other F-like ions considered in our earlier work \cite{ak1}, the 113 levels listed in the tables are lowest for Cs~XLVII alone. For other ions, some from other configurations, mainly 2s$^2$2p$^4$4$\ell$, intermix. To consider all listed levels of Tables~1--19 in strictly increasing order of energy, we have to consider up to 147 levels, although it varies from ion to ion. However, irrespective of the ion the first 94 levels are the {\em lowest} in energy.

In Table~C we compare our theoretical energies with GRASP and FAC codes, obtained with 501 and 38~089 levels, respectively, with the measurements of Hutton et al. \cite{rh} for the {\em common} levels of Ba~XLVIII, which have been recommended by the NIST team.  There are no large discrepancies for these levels, although differences are up to  $\sim$0.4~Ryd for  a few, such as 2s2p$^6$~$^2$S$_{1/2}$ (3),   2s$^2$2p$^4$($^3$P)3d ~$^2$P$  _{1/2}$ (9), 2s$^2$2p$^4$($^3$P)3d ~$^2$P$  _{3/2}$ (11), and 2s$^2$2p$^4$($^1$S)3d ~$^2$D$  _{5/2}$  (12). Furthermore, these differences are not systematic as for level 3 theoretical results are higher, but lower for others. In general, GRASP and FAC energies agree within 0.1~Ryd, the latter being mostly lower, and hence the agreement of NIST is slightly better with those with GRASP. We will also like to note here that the energy for the level 3 (2s2p$^6$~$^2$S$_{1/2}$) is not based on direct measurement, although J\"{o}nsson et al. \cite{jag} have been able to reproduce it with extensive CI, as discussed below.

In Table~D we compare our energies calculated with the GRASP and FAC codes with those of J\"{o}nsson et al. \cite{jag} for the two (common) levels 2s$^2$2p$^5$~$^2$P$^o_{1/2}$ and 2s2p$^6$~$^2$S$_{1/2}$, for all ions.  Also included in this table are the earlier energies of Gu \cite{gu}, obtained with the CI+MBPT code for the lowest six ions.  Differences between the GRASP and FAC calculations, or with the results of Gu,  are not significant and are within $\sim$0.2~Ryd for both levels. Similarly, there is a good agreement with the results of J\"{o}nsson et al. for the $^2$P$^o_{1/2}$ level, but discrepancies for $^2$S$_{1/2}$ are comparatively larger, and increase with increasing Z, as these are 0.2~Ryd for Cs~XLVII but 0.5~Ryd for Ta~LXV, with our results being invariably higher. Since the calculations of J\"{o}nsson et al. are comparatively more sophisticated, we believe these to be more accurate as well. Therefore, based on this comparison as well as the one discussed earlier, we may state that our energy levels are accurate to $\sim$0.5~Ryd. This conclusion is also based on comparisons between the GRASP and FAC energies  for a larger number of levels -- see Tables~1--19. However, we emphasize again that differences between the two calculations are not systematic, as for some the GRASP energies are higher whereas the reverse is true for the FAC -- see for example, levels 64 and 80 of Ba~XLVIII in Table~2. Therefore, inclusion of larger CI does not necessarily produce lower energies for all levels.

 \section{Radiative rates}\label{sec.eqs} 

Our calculated results  with the {\sc grasp} code are listed  in Tables 20--38 for transition energies (wavelengths, $\lambda_{ji}$ in ${\rm \AA}$), radiative rates (A-values, in s$^{-1}$), oscillator strengths (f-values, dimensionless), and line strengths (S-values, in atomic units, 1 a.u. = 6.460$\times$10$^{-36}$ cm$^2$ esu$^2$)  for E1 transitions in F-like ions with 55 $\le$ Z $\le$ 73.  These results are obtained in both the  velocity (Coulomb gauge) and length (Babushkin gauge) forms. Ideally both forms should give comparable results with their  ratio (R) close(r) to unity, but in practice they may differ substantially, especially for the very weak transitions with very small f-values. Nevertheless, we have also listed R in these tables for all E1 transitions, which are the most significant in any calculation.  Similar results for  E2, M1 and M2 transitions are listed only for the A-values, because the corresponding data for f- or S-values can be  obtained using Eqs. (1-5) given in \cite{kr}.   Finally, the {\em indices} used to represent the lower and upper levels of a transition are defined in Tables 1--19. Furthermore,  for brevity only transitions from the lowest 3 to higher excited levels are listed in Tables 20--38, but  full tables in the ASCII format are available online in the electronic version.

The only results for A-values available for the comparison purpose are those of Khatri et al. \cite{khat} for transitions from the lowest three to higher excited levels  of Ba~XLVIII. With them  we have no differences for the same reason as for the energy levels. However,  J\"{o}nsson et al. \cite{jag} have reported A-values for four transitions, namely  1--3 E1, 2--3 E1, 1--2 M1, and 1--2 E2, but for all F-like ions of interest. Therefore, in Table~E we make comparison between our and their results. It is highly satisfactory to note that there is no discrepancy for any transition and/or ion, and all results agree to better than 5\%. This is in spite of some differences in energies, particularly for the  2s2p$^6$~$^2$S$_{1/2}$ level, as seen earlier in Table~D. This confirms (yet again) that  small differences in transition energies do not lead to any appreciable differences in the subsequent results for A-values. In conclusion,  based on our experience and  comparisons made earlier for other F-like ions \cite{ak1}, our assessment of accuracy for the f- (and A-) values for a majority of strong transitions is $\sim$20\%, for all ions. 

Some times A-values for E3 (electric octupole) transitions may also be useful, if their strengths are comparable to others, such as the M2. However, for the F-like ions under consideration this is not the case. For example, for Cs~XLVII there are 2299 possible E3 transitions among the 113 levels, but only 10 of these have f $\sim$ 10$^{-6}$ and the rest are much weaker. On the other hand, there are 5 M2 transitions with f $\sim$ 10$^{-5}$, i.e. the E3 transitions are weaker than M2 by at least an order of magnitude. Similarly, for Gd~LVI there are 19 E3 and 10 M2 such transitions with f $\sim$ 10$^{-6}$ and f $\sim$ 10$^{-5}$, respectively. Finally, for Ta~LXV there are 19 E3 and 13 M2 such transitions with f $\sim$ 10$^{-6}$ and f $\sim$ 10$^{-5}$, respectively. Therefore, we have not included A-values for E3 transitions in Tables~20--38 but the results can be obtained from the author on request.

\section{Lifetimes}

The lifetime $\tau$ of a level $j$ is related to the A-values  as  1.0/$\Sigma_{i}$A$_{ji}$. As stated earlier, E1 transitions are (normally) the most dominant, and hence important in the determination of $\tau$. However,  summation over all types of transitions, i.e. E1, E2, M1, and M2, improves the accuracy and is particularly important for those levels for which there are no (strong) E1 connections.  Although $\tau$ is a measurable quantity, no experiments have yet been performed for transitions/levels of F-like ions of present interest. Therefore, no hard assessments of accuracy can be made. Unfortunately, the situation is no better with the theory. However, Khatri et al. \cite{khat} have listed $\tau$ for the levels of Ba~XLVIII for which we have no differences, {\em except} for level 3, i.e. 2s2p$^6$~$^2$S$_{1/2}$. For this level their listed $\tau$ is 1.43$\times$10$^{-13}$~s, whereas our result is 4.11$\times$10$^{-13}$~s, larger by a factor of three. Their result for this level is {\em incorrect}, because the dominant contributing transition for this is 1--3~E1 for which their A-value is 2.34$\times$10$^{12}$~s$^{-1}$, which leads to $\tau$ = 4.27$\times$10$^{-13}$~s, closer to our calculation. A similar discrepancy was noted  \cite{kmq} in their results for the 2s$^2$2p$^5$~$^2$P$^o_{1/2}$ level of Sr~XXX. For future comparisons with experimental or theoretical data our calculated values of $\tau$ are included  in Tables 1--19.  Since these results are directly related to the A-values, our assessment of accuracy for these is also the same, i.e. $\sim$20\%.

\section{Conclusions}

In this paper, energies for 113 levels of the 2s$^2$2p$^5$, 2s2p$^6$, 2s$^2$2p$^4$3$\ell$, 2s2p$^5$3$\ell$, and 2p$^6$3$\ell$ configurations of  19 F-like ions with 55 $\le$ Z $\le$ 73 are reported. Combined with our earlier results \cite{kr,xe,ak1}, this presents a complete data for ions with Z $\le$ 74.  Similarly as earlier, we have adopted the {\sc grasp} code for the calculations.  Since no existing data are available for most of the levels and ions with which to make comparisons, we have made additional calculations with the {\sc fac} code, but with much more extensive CI. This helps in assessing the accuracy of the energy levels. Based on several calculations with both codes, as well as comparisons with available limited theoretical and experimental data, our energy levels are assessed to be accurate to better than 0.5\% (0.5~Ryd), for all ions. However, for a few levels of each ion there is some ambiguity in their designations. This is because of very strong mixing with  one eigenvector of a CSF often dominating in magnitude for several levels. For this reason, mixing coefficients are listed for the levels of Cs~XLVII, as an example. However, similar results for other ions can be obtained from the author on request. 

Radiative rates for four types of transitions, i.e. E1, E2, M1, and M2,  are also reported among the above listed 113 levels. These data are significantly more extensive than currently  available in the literature.  Comparisons with the existing literature are limited to only four transitions, for which there are no discrepancies. Similarly,  calculations for comparatively larger  number of transitions for Ba~XLVIII are also available.  Again, there are no discrepancies with our results because of the inclusion of similar level of CI and code.  Based on whatever comparisons are possible and with our past experience on calculations for a wide range of ions, we assess our A-values and lifetimes  to be  accurate to $\sim$20\%, particularly for  strong transitions with large f-values. For very weak transitions the reported A-values may be comparatively less reliable.

%You are welcome to use BiBTeX with the \adndtbst\ bibliography
%style distributed with \adndtstyle\ package. This style comes very close
%to the journal style. Be sure to provide your
%\texttt{.bbl}  file (not the \texttt{.bib} file) with your submission.

%Please see \adndtguide\ for more instructions.

%\ack
%KMA  is thankful to  AWE Aldermaston for financial support. 

\begin{appendix}

\def\thesection{} % To get the appendix heading correct

\section{Appendix A. Supplementary data}% The Appendix itself

Owing to space limitations, only parts of Tables 20-38  are presented here, but full tables are being made available as supplemental material in conjunction with the electronic publication of this work. Supplementary data associated with this article can be found, in the online version, at doi:nn.nnnn/j.adt.2018.nn.nnn.

\end{appendix}

%%  All sections inside the appendix environment will be appendixes
%%  Subsections function normally in appendixes.
%\newpage

%\end{document}

%\vspace*{1.1 cm}
\clearpage
\newpage
\renewcommand{\baselinestretch}{1.0}
\footnotesize
\begin{longtable}{@{\extracolsep\fill}rllllrrrrr@{}}
\caption{Ground level  (2s$^2$2p$^5$~$^2$P$^o_{3/2}$) energies  (in Ryd) from the GRASP calculations with contributions from Breit and QED effects for  F-like ions with 55 $\le$ Z $\le$ 73.}
Z  & Ion & Coulomb & Breit & QED & Total  \\  \\
\hline \\
\endfirsthead\\
\caption[]{(continued)}
Z  & Ion & Coulomb & Breit & QED & Total  \\  \\
\hline \\

\endhead 
55 & Cs~XLVII  & -10995.70  &  09.09   & 07.56 &  -10979.05  \\
56 & Ba~XLVIII & -11430.84  &  09.63   & 08.04 &  -11413.17  \\
57 & La~XLIX   & -11875.54  &  10.2    & 08.54 &  -11856.80  \\
58 & Ce~L      & -12329.87  &  10.8    & 09.06 &  -12310.03  \\
59 & Pr~LI     & -12793.94  &  11.4    & 09.60 &  -12772.95  \\
60 & Nd~LII    & -13267.86  &  12.0    & 10.2  &  -13245.67  \\
61 & Pm~LIII   & -13751.73  &  12.7    & 10.8  &  -13728.29  \\
62 & Sm~LIV    & -14245.65  &  13.4    & 11.4  &  -14220.92  \\
63 & Eu~LV     & -14749.75  &  14.1    & 12.0  &  -14723.66  \\
64 & Gd~LVI    & -15264.14  &  14.8    & 12.7  &  -15236.64  \\
65 & Tb~LVII   & -15788.93  &  15.6    & 13.4  &  -15759.98  \\
66 & Dy~LVIII  & -16324.27  &  16.3    & 14.1  &  -16293.80  \\
67 & Ho~LIX    & -16870.27  &  17.2    & 14.9  &  -16838.24  \\
68 & Er~LX     & -17427.08  &  18.0    & 15.7  &  -17393.41  \\
69 & Tm~LXI    & -17994.83  &  18.9    & 16.5  &  -17959.48  \\
70 & Yb~LXII   & -18573.68  &  19.8    & 17.3  &  -18536.58  \\
71 & Lu~LXIII  & -19163.78  &  20.7    & 18.2  &  -19124.86  \\
72 & Hf~LXIV   & -19765.28  &  21.7    & 19.1  &  -19724.48  \\
73 & Ta~LXV    & -20378.36  &  22.7    & 20.0  &  -20335.61  \\
\\  \hline            								                	 
\end{longtable}   								   					       
			      							   					       
%\vspace*{0.5 cm}													       

%}													       
														       
\begin{flushleft}													       
{\small
   
}															       
\end{flushleft} 

%\end{document}

\clearpage
\newpage

%The tables that are part of the introductory material should be located after the figures, one table per page.

\renewcommand{\baselinestretch}{1.0}
\footnotesize
\begin{longtable}{@{\extracolsep\fill}rllr@{}}
\caption{Mixing coefficients (MC) for 113 levels of Cs~XLVII. Numbers outside and inside a bracket correspond to MC and the level, \\respectively.} %See Table~1 for the definition of all levels.
Index  & Configuration           & Level              &  Mixing coefficients     \\  \\
 \hline\\
\endfirsthead\\
\caption[]{(continued)}
Index  & Configuration           & Level              &  Mixing coefficients   \\  \\
\hline\\
\endhead
    1    &   2s$^2$2p$^5$	     &    $^2$P$^o_{3/2}$ &  1.00(  1)                                                                          \\
    2    &   2s$^2$2p$^5$	     &    $^2$P$^o_{1/2}$ &  1.00(  2)									        \\
    3    &   2s2p$^6$		     &    $^2$S$  _{1/2}$ &  1.00(  3)									        \\
    4    &   2s$^2$2p$^4$3s	     &    $^4$P$  _{5/2}$ &  0.84(  4)$+$0.54( 28)							        \\
    5    &   2s$^2$2p$^4$3s	     &    $^2$P$  _{3/2}$ &  0.35( 26)$+$0.75(  5)$+$0.55( 29)						        \\
    6    &   2s$^2$2p$^4$3s	     &    $^2$S$  _{1/2}$ &  0.51( 62)$-$0.34( 27)$+$0.79(  6)						        \\
    7    &   2s$^2$2p$^4$($^3$P)3p   &    $^4$P$^o_{3/2}$ & $-$0.27( 31)$+$0.58(  7)$-$0.43( 13)$+$0.30( 14)$+$0.37( 37)$-$0.40( 33)	        \\
    8    &   2s$^2$2p$^4$($^3$P)3p   &    $^2$D$^o_{5/2}$ &  0.53( 34)$-$0.36( 10)$+$0.54(  8)$+$0.48( 32)$-$0.26( 39)			        \\
\hline  											       								   					      
\end{longtable}

%\vspace*{0.5 cm}                                                                               
                               
\begin{flushleft}                                                                               
                               
{\small
                                                                               
}                                                                                               
                               
\end{flushleft} 
%}                        
%\end{document}

\clearpage
\newpage
\renewcommand{\baselinestretch}{1.0}
\footnotesize
\begin{longtable}{@{\extracolsep\fill}rllrrrr@{}}
\caption{Comparison of energy levels (in Ryd) for F-like Ba~XLVIII.}
S.No. & Configuration & Level & Index & NIST & GRASP & FAC  \\  \\
 \hline\\
\endfirsthead\\
\caption[]{(continued)}
S.No. & Configuration & Level & Index & NIST & GRASP & FAC  \\  \\
\hline\\
\endhead 
 1 & 2s$^2$2p$^5$	   & $^2$P$^o_{3/2}$ &     1  &   0.000   &   0.0000  &   0.0000  \\
 2 & 2s$^2$2p$^5$	   & $^2$P$^o_{1/2}$ &     2  &  28.985   &  28.9304  &  28.9524  \\
 3 & 2s2p$^6$		   & $^2$S$  _{1/2}$ &     3  &  52.450   &  52.6383  &  52.6054  \\
 4 & 2s$^2$2p$^4$3s	   & $^4$P$  _{5/2}$ &     4  &  343.793  & 343.7281  & 343.7235  \\
 5 & 2s$^2$2p$^4$3s	   & $^2$P$  _{3/2}$ &     5  &  344.366  & 344.3027  & 344.2899  \\
 6 & 2s$^2$2p$^4$3s	   & $^2$S$  _{1/2}$ &     6  &  346.930  & 346.8856  & 346.9050  \\
 7 & 2s$^2$2p$^4$($^3$P)3p & $^2$D$^o_{5/2}$ &     8  &  350.487  & 350.5544  & 350.5334  \\
 8 & 2s$^2$2p$^4$($^3$P)3p & $^4$S$^o_{3/2}$ &    13  &  360.862  & 360.7613  & 360.7373  \\
\hline            								                	 
\end{longtable}   								   					       
			      							   					       
%\vspace*{0.5 cm}													       

%}													       
														       
\begin{flushleft}													       
{\small
Index: see Table~2 \\
NIST: {\tt http://www.nist.gov/pml/data/asd.cfm}, energies for the lowest 3 levels are extrapolated results from Feldman et al. \cite{uf2} whereas for the
remaining levels are from measurements of Hutton et al. \cite{rh} \\
%Experimental: see  J\"{o}nsson  et al. \cite{} \\
GRASP: present calculations with  the {\sc grasp} code for 501 levels \\
FAC: present calculations with  the {\sc fac} code for 38~089 levels \\       
}															       
\end{flushleft} 

%\end{document}

\renewcommand{\baselinestretch}{1.0}
\footnotesize
\begin{longtable}{@{\extracolsep\fill}rrrrrrrrrr@{}}
\caption{Comparison of energies  (in Ryd) for the 2s$^2$2p$^5$~$^2$P$^o_{1/2}$ and 2s2p$^6$~$^2$S$_{1/2}$ levels of F-like ions with 55 $\le$ Z $\le$ 73.}
Z  & \multicolumn{2}{c}{J\"{o}nsson et al. \cite{jag}}    & \multicolumn{2}{c}{GRASP}      &   \multicolumn{2}{c}{FAC}   &   \multicolumn{2}{c}{MBPT} \\  \\
\hline \\
Level & $^2$P$^o_{1/2}$ & $^2$S$_{1/2}$  & $^2$P$^o_{1/2}$ & $^2$S$_{1/2}$  & $^2$P$^o_{1/2}$ & $^2$S$_{1/2}$   & $^2$P$^o_{1/2}$ & $^2$S$_{1/2}$  \\
\\ \hline  \\  
\endfirsthead\\
\caption[]{(continued)}
Z  & \multicolumn{2}{c}{J\"{o}nsson et al. \cite{jag}}    & \multicolumn{2}{c}{GRASP}      &   \multicolumn{2}{c}{FAC}   &   \multicolumn{2}{c}{MBPT} \\  \\
\hline \\
Level & $^2$P$^o_{1/2}$ & $^2$S$_{1/2}$   & $^2$P$^o_{1/2}$ & $^2$S$_{1/2}$  & $^2$P$^o_{1/2}$ & $^2$S$_{1/2}$   & $^2$P$^o_{1/2}$ & $^2$S$_{1/2}$  \\
\\ \hline  \\  \\
\hline\\
\endhead 
55 & 26.7247 & 49.6325    &  26.6976 & 49.8336  & 26.7173  & 49.8034	& 26.7550   & 49.6744 \\
56 & 28.9589 & 52.4309    &  28.9304 & 52.6383  & 28.9524  & 52.6054	& 28.9923   & 52.4764 \\
57 & 31.3379 & 55.3814    &  31.3081 & 55.5955  & 31.3325  & 55.5597	& 31.3747   & 55.4309 \\
58 & 33.8689 & 58.4913    &  33.8377 & 58.7126  & 33.8648  & 58.6735	& 33.9093   & 58.5451 \\
59 & 36.5593 & 61.7680    &  36.5267 & 61.9972  & 36.5566  & 61.9544	& 36.6037   & 61.8263 \\
60 & 39.4168 & 65.1892    &  39.3827 & 65.4574  & 39.4158  & 65.4098	& 39.4653   & 65.2821 \\
61 & 42.4496 & 68.8538    &  42.4138 & 69.1014  & 42.4502  & 69.0482	&	   &	     \\
62 & 45.6657 & 72.6734    &  45.6283 & 72.9380  & 45.6682  & 72.8783	&	   &	     \\
\hline            								                	 
\end{longtable}   								   					       
			      							   					       
%\vspace*{0.5 cm}													       

%}													       
														       
\begin{flushleft}													       
{\small
GRASP: present calculations with  the {\sc grasp} code for 501 levels \\
FAC: present calculations with  the {\sc fac} code for 38~089 levels \\ 
MBPT: earlier calculations of Gu \cite{gu} with  the {\sc MBPT} code  \\   
}															       
\end{flushleft} 

%\end{document}
\clearpage
\newpage

\renewcommand{\baselinestretch}{1.0}
\footnotesize
\begin{longtable}{@{\extracolsep\fill}rrrrr@{}}
\caption{Comparison of A-values (s$^{-1}$) for transitions among the lowest three levels of F-like ions with 55 $\le$ Z $\le$ 73. The first entry is from the present calculations with {\sc grasp} and the second is from J\"{o}nsson  et al. \cite{jag}.   $a{\pm}b \equiv a{\times}$10$^{{\pm}b}$.}
Z  & 1--3 (E1)& 2--3 (E1)& 1--2 (M1)& 1--2 (E2)  \\ \\
 \hline\\
\endfirsthead\\
\caption[]{(continued)}
Z  & 1--3 (E1)& 2--3 (E1)& 1--2 (M1)& 1--2 (E2) \\  \\
\hline\\
\endhead
55 & 2.064+12  &  9.531+10  &  4.418+08  &  1.112+06  \\
55 & 2.017+12  &  9.149+10  &  4.433+08  &  1.109+06  \\
56 & 2.337+12  &  9.821+10  &  5.616+08  &  1.526+06  \\
56 & 2.285+12  &  9.427+10  &  5.634+08  &  1.522+06  \\
57 & 2.646+12  &  1.012+11  &  7.109+08  &  2.084+06  \\
57 & 2.589+12  &  9.711+10  &  7.132+08  &  2.079+06  \\
58 & 2.998+12  &  1.042+11  &  8.965+08  &  2.832+06  \\
58 & 2.935+12  &  1.000+11  &  8.992+08  &  2.824+06  \\
\hline  											       								   					      
\end{longtable}

%\vspace*{0.5 cm}                                                                               
                               
\begin{flushleft}                                                                               
                               
{\small
                                                                            
}                                                                                               
                               
\end{flushleft} 
%}                        
%\end{document}

\clearpage
\newpage

%%% Please start a new page by uncommenting the next

\TableExplanation

\bigskip
\renewcommand{\arraystretch}{1.0}

\section*{Table 1.\label{tbl1te} Energies (Ryd) for 113 levels of Cs~XLVII and their lifetimes ($\tau$, s). For the ground level the energy is absolute whereas for others  are comparative.}
\begin{tabular}{@{}p{1in}p{6in}@{}}
Index            & Level Index \\
Configuration    & The configuration to which the level belongs \\
Level             & The $LSJ$ designation of the level \\
GRASP          & Present energies from the {\sc grasp} code  with 38  configurations and 501 level calculations \\
FAC               & Present energies from the {\sc fac} code  with 38~089 level calculations \\
$\tau$ (s)       & Lifetime of the level in s with the GRASP code \\

\end{tabular}
\label{tableII}

\bigskip
\renewcommand{\arraystretch}{1.0}

\section*{Table 2.\label{tbl2te} Energies (Ryd) for 113 levels of Ba~XLVIII and their lifetimes ($\tau$, s). For the ground level the energy is absolute whereas for others  are comparative.}
\begin{tabular}{@{}p{1in}p{6in}@{}}
Index            & Level Index \\
Configuration    & The configuration to which the level belongs \\
Level             & The $LSJ$ designation of the level \\
NIST              & Energies compiled by NIST and available at the website  {\tt http://www.nist.gov/pml/data/asd.cfm} \\
GRASP          & Present energies from the {\sc grasp} code  with 38  configurations and 501 level calculations \\
FAC1             & Present energies from the {\sc fac} code  with 501 level calculations \\
FAC2            & Present energies from the {\sc fac} code  with 38~089 level calculations \\
$\tau$ (s)       & Lifetime of the level in s \\

\end{tabular}
\label{tableII}

\bigskip
\renewcommand{\arraystretch}{1.0}

\section*{Table 3.\label{tbl3te} Energies (Ryd) for  113 levels of La~XLIX and their lifetimes ($\tau$, s). For the ground level the energy is absolute whereas for others  are comparative.}
\begin{tabular}{@{}p{1in}p{6in}@{}}
Index            & Level Index \\
Configuration    & The configuration to which the level belongs \\
Level             & The $LSJ$ designation of the level \\
GRASP          & Present energies from the {\sc grasp} code  with 38  configurations and 501 level calculations \\
FAC               & Present energies from the {\sc fac} code  with 38~089 level calculations \\
$\tau$ (s)       & Lifetime of the level in s with the GRASP code \\

\end{tabular}
\label{tableII}

\bigskip
\renewcommand{\arraystretch}{1.0}

\section*{Table 4.\label{tbl4te} Energies (Ryd) for 113 levels of Ce~L and their lifetimes ($\tau$, s). For the ground level the energy is absolute whereas for others  are comparative.}
\begin{tabular}{@{}p{1in}p{6in}@{}}
Index            & Level Index \\
Configuration    & The configuration to which the level belongs \\
Level             & The $LSJ$ designation of the level \\
GRASP          & Present energies from the {\sc grasp} code  with 38  configurations and 501 level calculations \\
FAC               & Present energies from the {\sc fac} code  with 38~089 level calculations \\
$\tau$ (s)       & Lifetime of the level in s with the GRASP code \\

\end{tabular}
\label{tableII}

\bigskip
\renewcommand{\arraystretch}{1.0}

\section*{Table 5.\label{tbl5te} Energies (Ryd) for 113 levels of Pr~LI and their lifetimes ($\tau$, s). For the ground level the energy is absolute whereas for others  are comparative.}
\begin{tabular}{@{}p{1in}p{6in}@{}}
Index            & Level Index \\
Configuration    & The configuration to which the level belongs \\
Level             & The $LSJ$ designation of the level \\
GRASP          & Present energies from the {\sc grasp} code  with 38  configurations and 501 level calculations \\
FAC               & Present energies from the {\sc fac} code  with 38~089 level calculations \\
$\tau$ (s)       & Lifetime of the level in s with the GRASP code \\

\end{tabular}
\label{tableII}

\bigskip
\renewcommand{\arraystretch}{1.0}

\section*{Table 6.\label{tbl6te} Energies (Ryd) for  113  levels of Nd~LII and their lifetimes ($\tau$, s). For the ground level the energy is absolute whereas for others  are comparative.}
\begin{tabular}{@{}p{1in}p{6in}@{}}
Index            & Level Index \\
Configuration    & The configuration to which the level belongs \\
Level             & The $LSJ$ designation of the level \\
GRASP          & Present energies from the {\sc grasp} code  with 38  configurations and 501 level calculations \\
FAC               & Present energies from the {\sc fac} code  with 38~089 level calculations \\
$\tau$ (s)       & Lifetime of the level in s with the GRASP code \\
\end{tabular}
\label{tableII}

\bigskip
\renewcommand{\arraystretch}{1.0}

\section*{Table 7.\label{tbl7te} Energies (Ryd) for 113 levels of Pm~LIII and their lifetimes ($\tau$, s). For the ground level the energy is absolute whereas for others  are comparative.}
\begin{tabular}{@{}p{1in}p{6in}@{}}
Index            & Level Index \\
Configuration    & The configuration to which the level belongs \\
Level             & The $LSJ$ designation of the level \\
GRASP          & Present energies from the {\sc grasp} code  with 38  configurations and 501 level calculations \\
FAC               & Present energies from the {\sc fac} code  with 38~089 level calculations \\
$\tau$ (s)       & Lifetime of the level in s with the GRASP code \\

\end{tabular}
\label{tableII}

\bigskip
\renewcommand{\arraystretch}{1.0}

\section*{Table 8.\label{tbl8te} Energies (Ryd) for  113 levels of Sm~LIV and their lifetimes ($\tau$, s). For the ground level the energy is absolute whereas for others  are comparative.}
\begin{tabular}{@{}p{1in}p{6in}@{}}
Index            & Level Index \\
Configuration    & The configuration to which the level belongs \\
Level             & The $LSJ$ designation of the level \\
GRASP          & Present energies from the {\sc grasp} code  with 38  configurations and 501 level calculations \\
FAC               & Present energies from the {\sc fac} code  with 38~089 level calculations \\
$\tau$ (s)       & Lifetime of the level in s with the GRASP code \\

\end{tabular}
\label{tableII}

\bigskip
\renewcommand{\arraystretch}{1.0}

\section*{Table 9.\label{tbl9te} Energies (Ryd) for  113 levels of Eu~LV and their lifetimes ($\tau$, s). For the ground level the energy is absolute whereas for others  are comparative.}
\begin{tabular}{@{}p{1in}p{6in}@{}}
Index            & Level Index \\
Configuration    & The configuration to which the level belongs \\
Level             & The $LSJ$ designation of the level \\
GRASP          & Present energies from the {\sc grasp} code  with 38  configurations and 501 level calculations \\
FAC               & Present energies from the {\sc fac} code  with 38~089 level calculations \\
$\tau$ (s)       & Lifetime of the level in s with the GRASP code \\

\end{tabular}
\label{tableII}

\bigskip
\renewcommand{\arraystretch}{1.0}

\section*{Table 10.\label{tbl10te} Energies (Ryd) for 113 levels of Gd~LVI and their lifetimes ($\tau$, s). For the ground level the energy is absolute whereas for others  are comparative.}
\begin{tabular}{@{}p{1in}p{6in}@{}}
Index            & Level Index \\
Configuration    & The configuration to which the level belongs \\
Level             & The $LSJ$ designation of the level \\
GRASP          & Present energies from the {\sc grasp} code  with 38  configurations and 501 level calculations \\
FAC               & Present energies from the {\sc fac} code  with 38~089 level calculations \\
$\tau$ (s)       & Lifetime of the level in s with the GRASP code \\
\end{tabular}
\label{tableII}

\bigskip
\renewcommand{\arraystretch}{1.0}

\section*{Table 11.\label{tbl11te} Energies (Ryd) for  113 levels of Tb~LVII and their lifetimes ($\tau$, s).  For the ground level the energy is absolute whereas for others  are comparative.}
\begin{tabular}{@{}p{1in}p{6in}@{}}
Index            & Level Index \\
Configuration    & The configuration to which the level belongs \\
Level             & The $LSJ$ designation of the level \\
GRASP          & Present energies from the {\sc grasp} code  with 38  configurations and 501 level calculations \\
FAC               & Present energies from the {\sc fac} code  with 38~089 level calculations \\
$\tau$ (s)       & Lifetime of the level in s with the GRASP code \\
\end{tabular}
\label{tableII}

\bigskip
\renewcommand{\arraystretch}{1.0}

\section*{Table 12.\label{tbl12te} Energies (Ryd) for 113 levels of Dy~LVIII and their lifetimes ($\tau$, s).  For the ground level the energy is absolute whereas for others  are comparative.}
\begin{tabular}{@{}p{1in}p{6in}@{}}
Index            & Level Index \\
Configuration    & The configuration to which the level belongs \\
Level             & The $LSJ$ designation of the level \\
GRASP          & Present energies from the {\sc grasp} code  with 38  configurations and 501 level calculations \\
FAC               & Present energies from the {\sc fac} code  with 38~089 level calculations \\
$\tau$ (s)       & Lifetime of the level in s with the GRASP code \\

\end{tabular}
\label{tableII}

\bigskip
\renewcommand{\arraystretch}{1.0}

\section*{Table 13.\label{tbl13te} Energies (Ryd) for  113 levels of Ho~LIX and their lifetimes ($\tau$, s).  For the ground level the energy is absolute whereas for others  are comparative.}
\begin{tabular}{@{}p{1in}p{6in}@{}}
Index            & Level Index \\
Configuration    & The configuration to which the level belongs \\
Level             & The $LSJ$ designation of the level \\
GRASP          & Present energies from the {\sc grasp} code  with 38  configurations and 501 level calculations \\
FAC               & Present energies from the {\sc fac} code  with 38~089 level calculations \\
$\tau$ (s)       & Lifetime of the level in s with the GRASP code \\

\end{tabular}
\label{tableII}

\bigskip
\renewcommand{\arraystretch}{1.0}

\section*{Table 14.\label{tbl14te} Energies (Ryd) for  113  levels of Er~LX and their lifetimes ($\tau$, s).  For the ground level the energy is absolute whereas for others  are comparative.}
\begin{tabular}{@{}p{1in}p{6in}@{}}
Index            & Level Index \\
Configuration    & The configuration to which the level belongs \\
Level             & The $LSJ$ designation of the level \\
GRASP          & Present energies from the {\sc grasp} code  with 38  configurations and 501 level calculations \\
FAC               & Present energies from the {\sc fac} code  with 38~089 level calculations \\
$\tau$ (s)       & Lifetime of the level in s with the GRASP code \\
\end{tabular}
\label{tableII}

\bigskip
\renewcommand{\arraystretch}{1.0}

\section*{Table 15.\label{tbl15te} Energies (Ryd) for  113 levels of Tm~LXI and their lifetimes ($\tau$, s).   For the ground level the energy is absolute whereas for others  are comparative.}
\begin{tabular}{@{}p{1in}p{6in}@{}}
Index            & Level Index \\
Configuration    & The configuration to which the level belongs \\
Level             & The $LSJ$ designation of the level \\
GRASP          & Present energies from the {\sc grasp} code  with 38  configurations and 501 level calculations \\
FAC               & Present energies from the {\sc fac} code  with 38~089 level calculations \\
$\tau$ (s)       & Lifetime of the level in s with the GRASP code \\

\end{tabular}
\label{tableII}

\bigskip
\renewcommand{\arraystretch}{1.0}

\section*{Table 16.\label{tbl16te} Energies (Ryd) for 113 levels of Yb~LXII and their lifetimes ($\tau$, s).  For the ground level the energy is absolute whereas for others  are comparative.}
\begin{tabular}{@{}p{1in}p{6in}@{}}
Index            & Level Index \\
Configuration    & The configuration to which the level belongs \\
Level             & The $LSJ$ designation of the level \\
GRASP          & Present energies from the {\sc grasp} code  with 38  configurations and 501 level calculations \\
FAC               & Present energies from the {\sc fac} code  with 38~089 level calculations \\
$\tau$ (s)       & Lifetime of the level in s with the GRASP code \\

\end{tabular}
\label{tableII}

\bigskip
\renewcommand{\arraystretch}{1.0}

\section*{Table 17.\label{tbl17te} Energies (Ryd) for  113 levels of Lu~LXIII and their lifetimes ($\tau$, s).  For the ground level the energy is absolute whereas for others  are comparative.}
\begin{tabular}{@{}p{1in}p{6in}@{}}
Index            & Level Index \\
Configuration    & The configuration to which the level belongs \\
Level             & The $LSJ$ designation of the level \\
GRASP          & Present energies from the {\sc grasp} code  with 38  configurations and 501 level calculations \\
FAC               & Present energies from the {\sc fac} code  with 38~089 level calculations \\
$\tau$ (s)       & Lifetime of the level in s with the GRASP code \\
\end{tabular}
\label{tableII}

\bigskip
\renewcommand{\arraystretch}{1.0}

\section*{Table 18.\label{tbl18te} Energies (Ryd) for  113 levels of Hf~LXIV and their lifetimes ($\tau$, s).  For the ground level the energy is absolute whereas for others  are comparative.}
\begin{tabular}{@{}p{1in}p{6in}@{}}
Index            & Level Index \\
Configuration    & The configuration to which the level belongs \\
Level             & The $LSJ$ designation of the level \\
GRASP          & Present energies from the {\sc grasp} code  with 38  configurations and 501 level calculations \\
FAC               & Present energies from the {\sc fac} code  with 38~089 level calculations \\
$\tau$ (s)       & Lifetime of the level in s with the GRASP code \\

\end{tabular}
\label{tableII}

\bigskip
\renewcommand{\arraystretch}{1.0}

\section*{Table 19.\label{tbl19te} Energies (Ryd) for  113 levels of Ta~LXV and their lifetimes ($\tau$, s).  For the ground level the energy is absolute whereas for others  are comparative.}
\begin{tabular}{@{}p{1in}p{6in}@{}}
Index            & Level Index \\
Configuration    & The configuration to which the level belongs \\
Level             & The $LSJ$ designation of the level \\
GRASP          & Present energies from the {\sc grasp} code  with 38  configurations and 501 level calculations \\
FAC               & Present energies from the {\sc fac} code  with 38~089 level calculations \\
$\tau$ (s)       & Lifetime of the level in s with the GRASP code \\

\end{tabular}
\label{tableII}

%The following serves as an example on how to format a complicated
%explanation of table. The corresponding Table is however missing in this
%template.
%\end{document}

\bigskip
\section*{Table 20.\label{tbl20te}  Transition wavelengths ($\lambda_{ij}$ in $\rm \AA$), radiative rates (A$_{ji}$ in s$^{-1}$),
 oscillator strengths (f$_{ij}$, dimensionless), and line strengths (S, in atomic units) for electric dipole (E1), and 
A$_{ji}$ for electric quadrupole (E2), magnetic dipole (M1), and magnetic quadrupole (M2) transitions of Cs~XLVII.
 The ratio R(E1) of velocity and length forms of A-values for E1 transitions is listed in the last column.}
\begin{tabular}{@{}p{1in}p{6in}@{}}
$i$ and $j$         & The lower ($i$) and upper ($j$) levels of a transition as defined in Table 1.\\
$\lambda_{ij}$      & Transition wavelength (in ${\rm \AA}$) \\
A$^{E1}_{ji}$       & Radiative transition probability (in s$^{-1}$) for the E1 transitions \\
f$^{E1}_{ij}$       & Absorption oscillator strength (dimensionless) for the E1 transitions \\
S$^{E1}$            & Line strength in atomic unit (a.u.), 1 a.u. = 6.460$\times$10$^{-36}$ cm$^2$ esu$^2$ for the E1 transitions \\
A$^{E2}_{ji}$       & Radiative transition probability (in s$^{-1}$) for the E2 transitions \\
A$^{M1}_{ji}$       & Radiative transition probability (in s$^{-1}$) for the M1 transitions \\
A$^{M2}_{ji}$       & Radiative transition probability (in s$^{-1}$) for the M2 transitions \\
R(E1)                     & Ratio of velocity and length forms of A- (or f- and S-) values for the E1 transitions \\
$a{\pm}b$ &  $\equiv a\times{10^{{\pm}b}}$ \\
\end{tabular}
\label{ExplTable20}

\bigskip
\section*{Table 21.\label{tbl21te}  Transition wavelengths ($\lambda_{ij}$ in $\rm \AA$), radiative rates (A$_{ji}$ in s$^{-1}$),
 oscillator strengths (f$_{ij}$, dimensionless), and line strengths (S, in atomic units) for electric dipole (E1), and 
A$_{ji}$ for electric quadrupole (E2), magnetic dipole (M1), and magnetic quadrupole (M2) transitions of Ba~XLVIII.
The ratio R(E1) of velocity and length forms of A-values for E1 transitions is listed in the last column.}
\begin{tabular}{@{}p{1in}p{6in}@{}}
$i$ and $j$         & The lower ($i$) and upper ($j$) levels of a transition as defined in Table 2.\\
$\lambda_{ij}$      & Transition wavelength (in ${\rm \AA}$) \\
A$^{E1}_{ji}$       & Radiative transition probability (in s$^{-1}$) for the E1 transitions \\
f$^{E1}_{ij}$       & Absorption oscillator strength (dimensionless) for the E1 transitions \\
S$^{E1}$            & Line strength in atomic unit (a.u.), 1 a.u. = 6.460$\times$10$^{-36}$ cm$^2$ esu$^2$ for the E1 transitions \\
A$^{E2}_{ji}$       & Radiative transition probability (in s$^{-1}$) for the E2 transitions \\
A$^{M1}_{ji}$       & Radiative transition probability (in s$^{-1}$) for the M1 transitions \\
A$^{M2}_{ji}$       & Radiative transition probability (in s$^{-1}$) for the M2 transitions \\
R(E1)                     & Ratio of velocity and length forms of A- (or f- and S-) values for the E1 transitions \\
$a{\pm}b$ &  $\equiv a\times{10^{{\pm}b}}$ \\
\end{tabular}
\label{ExplTable21}

\bigskip
\section*{Table 22.\label{tbl22te}  Transition wavelengths ($\lambda_{ij}$ in $\rm \AA$), radiative rates (A$_{ji}$ in s$^{-1}$),
 oscillator strengths (f$_{ij}$, dimensionless), and line strengths (S, in atomic units) for electric dipole (E1), and 
A$_{ji}$ for electric quadrupole (E2), magnetic dipole (M1), and magnetic quadrupole (M2) transitions of La~XLIX.
 The ratio R(E1) of velocity and length forms of A-values for E1 transitions is listed in the last column.}
\begin{tabular}{@{}p{1in}p{6in}@{}}
$i$ and $j$         & The lower ($i$) and upper ($j$) levels of a transition as defined in Table 3.\\
$\lambda_{ij}$      & Transition wavelength (in ${\rm \AA}$) \\
A$^{E1}_{ji}$       & Radiative transition probability (in s$^{-1}$) for the E1 transitions \\
f$^{E1}_{ij}$       & Absorption oscillator strength (dimensionless) for the E1 transitions \\
S$^{E1}$            & Line strength in atomic unit (a.u.), 1 a.u. = 6.460$\times$10$^{-36}$ cm$^2$ esu$^2$ for the E1 transitions \\
A$^{E2}_{ji}$       & Radiative transition probability (in s$^{-1}$) for the E2 transitions \\
A$^{M1}_{ji}$       & Radiative transition probability (in s$^{-1}$) for the M1 transitions \\
A$^{M2}_{ji}$       & Radiative transition probability (in s$^{-1}$) for the M2 transitions \\
R(E1)                     & Ratio of velocity and length forms of A- (or f- and S-) values for the E1 transitions \\
$a{\pm}b$ &  $\equiv a\times{10^{{\pm}b}}$ \\
\end{tabular}
\label{ExplTable22}

\bigskip
\section*{Table 23.\label{tbl23te}  Transition wavelengths ($\lambda_{ij}$ in $\rm \AA$), radiative rates (A$_{ji}$ in s$^{-1}$),
 oscillator strengths (f$_{ij}$, dimensionless), and line strengths (S, in atomic units) for electric dipole (E1), and 
A$_{ji}$ for electric quadrupole (E2), magnetic dipole (M1), and magnetic quadrupole (M2) transitions of Ce~L.
 The ratio R(E1) of velocity and length forms of A-values for E1 transitions is listed in the last column.}
\begin{tabular}{@{}p{1in}p{6in}@{}}
$i$ and $j$         & The lower ($i$) and upper ($j$) levels of a transition as defined in Table 4.\\
$\lambda_{ij}$      & Transition wavelength (in ${\rm \AA}$) \\
A$^{E1}_{ji}$       & Radiative transition probability (in s$^{-1}$) for the E1 transitions \\
f$^{E1}_{ij}$       & Absorption oscillator strength (dimensionless) for the E1 transitions \\
S$^{E1}$            & Line strength in atomic unit (a.u.), 1 a.u. = 6.460$\times$10$^{-36}$ cm$^2$ esu$^2$ for the E1 transitions \\
A$^{E2}_{ji}$       & Radiative transition probability (in s$^{-1}$) for the E2 transitions \\
A$^{M1}_{ji}$       & Radiative transition probability (in s$^{-1}$) for the M1 transitions \\
A$^{M2}_{ji}$       & Radiative transition probability (in s$^{-1}$) for the M2 transitions \\
R(E1)                     & Ratio of velocity and length forms of A- (or f- and S-) values for the E1 transitions \\
$a{\pm}b$ &  $\equiv a\times{10^{{\pm}b}}$ \\
\end{tabular}
\label{ExplTable23}

\bigskip
\section*{Table 24.\label{tbl24te}  Transition wavelengths ($\lambda_{ij}$ in $\rm \AA$), radiative rates (A$_{ji}$ in s$^{-1}$),
 oscillator strengths (f$_{ij}$, dimensionless), and line strengths (S, in atomic units) for electric dipole (E1), and 
A$_{ji}$ for electric quadrupole (E2), magnetic dipole (M1), and magnetic quadrupole (M2) transitions of Pr~LI.
 The ratio R(E1) of velocity and length forms of A-values for E1 transitions is listed in the last column.}
\begin{tabular}{@{}p{1in}p{6in}@{}}
$i$ and $j$         & The lower ($i$) and upper ($j$) levels of a transition as defined in Table 5.\\
$\lambda_{ij}$      & Transition wavelength (in ${\rm \AA}$) \\
A$^{E1}_{ji}$       & Radiative transition probability (in s$^{-1}$) for the E1 transitions \\
f$^{E1}_{ij}$       & Absorption oscillator strength (dimensionless) for the E1 transitions \\
S$^{E1}$            & Line strength in atomic unit (a.u.), 1 a.u. = 6.460$\times$10$^{-36}$ cm$^2$ esu$^2$ for the E1 transitions \\
A$^{E2}_{ji}$       & Radiative transition probability (in s$^{-1}$) for the E2 transitions \\
A$^{M1}_{ji}$       & Radiative transition probability (in s$^{-1}$) for the M1 transitions \\
A$^{M2}_{ji}$       & Radiative transition probability (in s$^{-1}$) for the M2 transitions \\
R(E1)                     & Ratio of velocity and length forms of A- (or f- and S-) values for the E1 transitions \\
$a{\pm}b$ &  $\equiv a\times{10^{{\pm}b}}$ \\
\end{tabular}
\label{ExplTable24}

\bigskip
\section*{Table 25.\label{tbl25te}  Transition wavelengths ($\lambda_{ij}$ in $\rm \AA$), radiative rates (A$_{ji}$ in s$^{-1}$),
 oscillator strengths (f$_{ij}$, dimensionless), and line strengths (S, in atomic units) for electric dipole (E1), and 
A$_{ji}$ for electric quadrupole (E2), magnetic dipole (M1), and magnetic quadrupole (M2) transitions of Nd~LII.
  The ratio R(E1) of velocity and length forms of A-values for E1 transitions is listed in the last column.}
\begin{tabular}{@{}p{1in}p{6in}@{}}
$i$ and $j$         & The lower ($i$) and upper ($j$) levels of a transition as defined in Table 6.\\
$\lambda_{ij}$      & Transition wavelength (in ${\rm \AA}$) \\
A$^{E1}_{ji}$       & Radiative transition probability (in s$^{-1}$) for the E1 transitions \\
f$^{E1}_{ij}$       & Absorption oscillator strength (dimensionless) for the E1 transitions \\
S$^{E1}$            & Line strength in atomic unit (a.u.), 1 a.u. = 6.460$\times$10$^{-36}$ cm$^2$ esu$^2$ for the E1 transitions \\
A$^{E2}_{ji}$       & Radiative transition probability (in s$^{-1}$) for the E2 transitions \\
A$^{M1}_{ji}$       & Radiative transition probability (in s$^{-1}$) for the M1 transitions \\
A$^{M2}_{ji}$       & Radiative transition probability (in s$^{-1}$) for the M2 transitions \\
R(E1)                     & Ratio of velocity and length forms of A- (or f- and S-) values for the E1 transitions \\
$a{\pm}b$ &  $\equiv a\times{10^{{\pm}b}}$ \\
\end{tabular}
\label{ExplTable25}

\bigskip
\section*{Table 26.\label{tbl26te}  Transition wavelengths ($\lambda_{ij}$ in $\rm \AA$), radiative rates (A$_{ji}$ in s$^{-1}$),
 oscillator strengths (f$_{ij}$, dimensionless), and line strengths (S, in atomic units) for electric dipole (E1), and 
A$_{ji}$ for electric quadrupole (E2), magnetic dipole (M1), and magnetic quadrupole (M2) transitions of Pm~LIII.
  The ratio R(E1) of velocity and length forms of A-values for E1 transitions is listed in the last column.}
\begin{tabular}{@{}p{1in}p{6in}@{}}
$i$ and $j$         & The lower ($i$) and upper ($j$) levels of a transition as defined in Table 7.\\
$\lambda_{ij}$      & Transition wavelength (in ${\rm \AA}$) \\
A$^{E1}_{ji}$       & Radiative transition probability (in s$^{-1}$) for the E1 transitions \\
f$^{E1}_{ij}$       & Absorption oscillator strength (dimensionless) for the E1 transitions \\
S$^{E1}$            & Line strength in atomic unit (a.u.), 1 a.u. = 6.460$\times$10$^{-36}$ cm$^2$ esu$^2$ for the E1 transitions \\
A$^{E2}_{ji}$       & Radiative transition probability (in s$^{-1}$) for the E2 transitions \\
A$^{M1}_{ji}$       & Radiative transition probability (in s$^{-1}$) for the M1 transitions \\
A$^{M2}_{ji}$       & Radiative transition probability (in s$^{-1}$) for the M2 transitions \\
R(E1)                     & Ratio of velocity and length forms of A- (or f- and S-) values for the E1 transitions \\
$a{\pm}b$ &  $\equiv a\times{10^{{\pm}b}}$ \\
\end{tabular}
\label{ExplTable26}

\bigskip
\section*{Table 27.\label{tbl27te}  Transition wavelengths ($\lambda_{ij}$ in $\rm \AA$), radiative rates (A$_{ji}$ in s$^{-1}$),
 oscillator strengths (f$_{ij}$, dimensionless), and line strengths (S, in atomic units) for electric dipole (E1), and 
A$_{ji}$ for electric quadrupole (E2), magnetic dipole (M1), and magnetic quadrupole (M2) transitions of Sm~LIV.
  The ratio R(E1) of velocity and length forms of A-values for E1 transitions is listed in the last column.}
\begin{tabular}{@{}p{1in}p{6in}@{}}
$i$ and $j$         & The lower ($i$) and upper ($j$) levels of a transition as defined in Table 8.\\
$\lambda_{ij}$      & Transition wavelength (in ${\rm \AA}$) \\
A$^{E1}_{ji}$       & Radiative transition probability (in s$^{-1}$) for the E1 transitions \\
f$^{E1}_{ij}$       & Absorption oscillator strength (dimensionless) for the E1 transitions \\
S$^{E1}$            & Line strength in atomic unit (a.u.), 1 a.u. = 6.460$\times$10$^{-36}$ cm$^2$ esu$^2$ for the E1 transitions \\
A$^{E2}_{ji}$       & Radiative transition probability (in s$^{-1}$) for the E2 transitions \\
A$^{M1}_{ji}$       & Radiative transition probability (in s$^{-1}$) for the M1 transitions \\
A$^{M2}_{ji}$       & Radiative transition probability (in s$^{-1}$) for the M2 transitions \\
R(E1)                     & Ratio of velocity and length forms of A- (or f- and S-) values for the E1 transitions \\
$a{\pm}b$ &  $\equiv a\times{10^{{\pm}b}}$ \\
\end{tabular}
\label{ExplTable27}

\bigskip
\section*{Table 28.\label{tbl28te}  Transition wavelengths ($\lambda_{ij}$ in $\rm \AA$), radiative rates (A$_{ji}$ in s$^{-1}$),
 oscillator strengths (f$_{ij}$, dimensionless), and line strengths (S, in atomic units) for electric dipole (E1), and 
A$_{ji}$ for electric quadrupole (E2), magnetic dipole (M1), and magnetic quadrupole (M2) transitions of Eu~LV.
 The ratio R(E1) of velocity and length forms of A-values for E1 transitions is listed in the last column.}
\begin{tabular}{@{}p{1in}p{6in}@{}}
$i$ and $j$         & The lower ($i$) and upper ($j$) levels of a transition as defined in Table 9.\\
$\lambda_{ij}$      & Transition wavelength (in ${\rm \AA}$) \\
A$^{E1}_{ji}$       & Radiative transition probability (in s$^{-1}$) for the E1 transitions \\
f$^{E1}_{ij}$       & Absorption oscillator strength (dimensionless) for the E1 transitions \\
S$^{E1}$            & Line strength in atomic unit (a.u.), 1 a.u. = 6.460$\times$10$^{-36}$ cm$^2$ esu$^2$ for the E1 transitions \\
A$^{E2}_{ji}$       & Radiative transition probability (in s$^{-1}$) for the E2 transitions \\
A$^{M1}_{ji}$       & Radiative transition probability (in s$^{-1}$) for the M1 transitions \\
A$^{M2}_{ji}$       & Radiative transition probability (in s$^{-1}$) for the M2 transitions \\
R(E1)                     & Ratio of velocity and length forms of A- (or f- and S-) values for the E1 transitions \\
$a{\pm}b$ &  $\equiv a\times{10^{{\pm}b}}$ \\
\end{tabular}
\label{ExplTable28}

\bigskip
\section*{Table 29.\label{tbl29te}  Transition wavelengths ($\lambda_{ij}$ in $\rm \AA$), radiative rates (A$_{ji}$ in s$^{-1}$),
 oscillator strengths (f$_{ij}$, dimensionless), and line strengths (S, in atomic units) for electric dipole (E1), and 
A$_{ji}$ for electric quadrupole (E2), magnetic dipole (M1), and magnetic quadrupole (M2) transitions of Gd~LVI.
The ratio R(E1) of velocity and length forms of A-values for E1 transitions is listed in the last column.}
\begin{tabular}{@{}p{1in}p{6in}@{}}
$i$ and $j$         & The lower ($i$) and upper ($j$) levels of a transition as defined in Table 10.\\
$\lambda_{ij}$      & Transition wavelength (in ${\rm \AA}$) \\
A$^{E1}_{ji}$       & Radiative transition probability (in s$^{-1}$) for the E1 transitions \\
f$^{E1}_{ij}$       & Absorption oscillator strength (dimensionless) for the E1 transitions \\
S$^{E1}$            & Line strength in atomic unit (a.u.), 1 a.u. = 6.460$\times$10$^{-36}$ cm$^2$ esu$^2$ for the E1 transitions \\
A$^{E2}_{ji}$       & Radiative transition probability (in s$^{-1}$) for the E2 transitions \\
A$^{M1}_{ji}$       & Radiative transition probability (in s$^{-1}$) for the M1 transitions \\
A$^{M2}_{ji}$       & Radiative transition probability (in s$^{-1}$) for the M2 transitions \\
R(E1)                     & Ratio of velocity and length forms of A- (or f- and S-) values for the E1 transitions \\
$a{\pm}b$ &  $\equiv a\times{10^{{\pm}b}}$ \\
\end{tabular}
\label{ExplTable29}

\bigskip
\section*{Table 30.\label{tbl30te}  Transition wavelengths ($\lambda_{ij}$ in $\rm \AA$), radiative rates (A$_{ji}$ in s$^{-1}$),
 oscillator strengths (f$_{ij}$, dimensionless), and line strengths (S, in atomic units) for electric dipole (E1), and 
A$_{ji}$ for electric quadrupole (E2), magnetic dipole (M1), and magnetic quadrupole (M2) transitions of Tb~LVII.
 The ratio R(E1) of velocity and length forms of A-values for E1 transitions is listed in the last column.}
\begin{tabular}{@{}p{1in}p{6in}@{}}
$i$ and $j$         & The lower ($i$) and upper ($j$) levels of a transition as defined in Table 11.\\
$\lambda_{ij}$      & Transition wavelength (in ${\rm \AA}$) \\
A$^{E1}_{ji}$       & Radiative transition probability (in s$^{-1}$) for the E1 transitions \\
f$^{E1}_{ij}$       & Absorption oscillator strength (dimensionless) for the E1 transitions \\
S$^{E1}$            & Line strength in atomic unit (a.u.), 1 a.u. = 6.460$\times$10$^{-36}$ cm$^2$ esu$^2$ for the E1 transitions \\
A$^{E2}_{ji}$       & Radiative transition probability (in s$^{-1}$) for the E2 transitions \\
A$^{M1}_{ji}$       & Radiative transition probability (in s$^{-1}$) for the M1 transitions \\
A$^{M2}_{ji}$       & Radiative transition probability (in s$^{-1}$) for the M2 transitions \\
R(E1)                     & Ratio of velocity and length forms of A- (or f- and S-) values for the E1 transitions \\
$a{\pm}b$ &  $\equiv a\times{10^{{\pm}b}}$ \\
\end{tabular}
\label{ExplTable30}

\bigskip
\section*{Table 31.\label{tbl31te}  Transition wavelengths ($\lambda_{ij}$ in $\rm \AA$), radiative rates (A$_{ji}$ in s$^{-1}$),
 oscillator strengths (f$_{ij}$, dimensionless), and line strengths (S, in atomic units) for electric dipole (E1), and 
A$_{ji}$ for electric quadrupole (E2), magnetic dipole (M1), and magnetic quadrupole (M2) transitions of Dy~LVIII.
 The ratio R(E1) of velocity and length forms of A-values for E1 transitions is listed in the last column.}
\begin{tabular}{@{}p{1in}p{6in}@{}}
$i$ and $j$         & The lower ($i$) and upper ($j$) levels of a transition as defined in Table 12.\\
$\lambda_{ij}$      & Transition wavelength (in ${\rm \AA}$) \\
A$^{E1}_{ji}$       & Radiative transition probability (in s$^{-1}$) for the E1 transitions \\
f$^{E1}_{ij}$       & Absorption oscillator strength (dimensionless) for the E1 transitions \\
S$^{E1}$            & Line strength in atomic unit (a.u.), 1 a.u. = 6.460$\times$10$^{-36}$ cm$^2$ esu$^2$ for the E1 transitions \\
A$^{E2}_{ji}$       & Radiative transition probability (in s$^{-1}$) for the E2 transitions \\
A$^{M1}_{ji}$       & Radiative transition probability (in s$^{-1}$) for the M1 transitions \\
A$^{M2}_{ji}$       & Radiative transition probability (in s$^{-1}$) for the M2 transitions \\
R(E1)                     & Ratio of velocity and length forms of A- (or f- and S-) values for the E1 transitions \\
$a{\pm}b$ &  $\equiv a\times{10^{{\pm}b}}$ \\
\end{tabular}
\label{ExplTable31}

\bigskip
\section*{Table 32.\label{tbl32te}  Transition wavelengths ($\lambda_{ij}$ in $\rm \AA$), radiative rates (A$_{ji}$ in s$^{-1}$),
 oscillator strengths (f$_{ij}$, dimensionless), and line strengths (S, in atomic units) for electric dipole (E1), and 
A$_{ji}$ for electric quadrupole (E2), magnetic dipole (M1), and magnetic quadrupole (M2) transitions of Ho~LIX.
 The ratio R(E1) of velocity and length forms of A-values for E1 transitions is listed in the last column.}
\begin{tabular}{@{}p{1in}p{6in}@{}}
$i$ and $j$         & The lower ($i$) and upper ($j$) levels of a transition as defined in Table 13.\\
$\lambda_{ij}$      & Transition wavelength (in ${\rm \AA}$) \\
A$^{E1}_{ji}$       & Radiative transition probability (in s$^{-1}$) for the E1 transitions \\
f$^{E1}_{ij}$       & Absorption oscillator strength (dimensionless) for the E1 transitions \\
S$^{E1}$            & Line strength in atomic unit (a.u.), 1 a.u. = 6.460$\times$10$^{-36}$ cm$^2$ esu$^2$ for the E1 transitions \\
A$^{E2}_{ji}$       & Radiative transition probability (in s$^{-1}$) for the E2 transitions \\
A$^{M1}_{ji}$       & Radiative transition probability (in s$^{-1}$) for the M1 transitions \\
A$^{M2}_{ji}$       & Radiative transition probability (in s$^{-1}$) for the M2 transitions \\
R(E1)                     & Ratio of velocity and length forms of A- (or f- and S-) values for the E1 transitions \\
$a{\pm}b$ &  $\equiv a\times{10^{{\pm}b}}$ \\
\end{tabular}
\label{ExplTable32}

\bigskip
\section*{Table 33.\label{tbl33te}  Transition wavelengths ($\lambda_{ij}$ in $\rm \AA$), radiative rates (A$_{ji}$ in s$^{-1}$),
 oscillator strengths (f$_{ij}$, dimensionless), and line strengths (S, in atomic units) for electric dipole (E1), and 
A$_{ji}$ for electric quadrupole (E2), magnetic dipole (M1), and magnetic quadrupole (M2) transitions of Er~LX.
  The ratio R(E1) of velocity and length forms of A-values for E1 transitions is listed in the last column.}
\begin{tabular}{@{}p{1in}p{6in}@{}}
$i$ and $j$         & The lower ($i$) and upper ($j$) levels of a transition as defined in Table 14.\\
$\lambda_{ij}$      & Transition wavelength (in ${\rm \AA}$) \\
A$^{E1}_{ji}$       & Radiative transition probability (in s$^{-1}$) for the E1 transitions \\
f$^{E1}_{ij}$       & Absorption oscillator strength (dimensionless) for the E1 transitions \\
S$^{E1}$            & Line strength in atomic unit (a.u.), 1 a.u. = 6.460$\times$10$^{-36}$ cm$^2$ esu$^2$ for the E1 transitions \\
A$^{E2}_{ji}$       & Radiative transition probability (in s$^{-1}$) for the E2 transitions \\
A$^{M1}_{ji}$       & Radiative transition probability (in s$^{-1}$) for the M1 transitions \\
A$^{M2}_{ji}$       & Radiative transition probability (in s$^{-1}$) for the M2 transitions \\
R(E1)                     & Ratio of velocity and length forms of A- (or f- and S-) values for the E1 transitions \\
$a{\pm}b$ &  $\equiv a\times{10^{{\pm}b}}$ \\
\end{tabular}
\label{ExplTable33}

\bigskip
\section*{Table 34.\label{tbl34te}  Transition wavelengths ($\lambda_{ij}$ in $\rm \AA$), radiative rates (A$_{ji}$ in s$^{-1}$),
 oscillator strengths (f$_{ij}$, dimensionless), and line strengths (S, in atomic units) for electric dipole (E1), and 
A$_{ji}$ for electric quadrupole (E2), magnetic dipole (M1), and magnetic quadrupole (M2) transitions of Tm~LXI.
  The ratio R(E1) of velocity and length forms of A-values for E1 transitions is listed in the last column.}
\begin{tabular}{@{}p{1in}p{6in}@{}}
$i$ and $j$         & The lower ($i$) and upper ($j$) levels of a transition as defined in Table 15.\\
$\lambda_{ij}$      & Transition wavelength (in ${\rm \AA}$) \\
A$^{E1}_{ji}$       & Radiative transition probability (in s$^{-1}$) for the E1 transitions \\
f$^{E1}_{ij}$       & Absorption oscillator strength (dimensionless) for the E1 transitions \\
S$^{E1}$            & Line strength in atomic unit (a.u.), 1 a.u. = 6.460$\times$10$^{-36}$ cm$^2$ esu$^2$ for the E1 transitions \\
A$^{E2}_{ji}$       & Radiative transition probability (in s$^{-1}$) for the E2 transitions \\
A$^{M1}_{ji}$       & Radiative transition probability (in s$^{-1}$) for the M1 transitions \\
A$^{M2}_{ji}$       & Radiative transition probability (in s$^{-1}$) for the M2 transitions \\
R(E1)                     & Ratio of velocity and length forms of A- (or f- and S-) values for the E1 transitions \\
$a{\pm}b$ &  $\equiv a\times{10^{{\pm}b}}$ \\
\end{tabular}
\label{ExplTable34}

\bigskip
\section*{Table 35.\label{tbl35te}  Transition wavelengths ($\lambda_{ij}$ in $\rm \AA$), radiative rates (A$_{ji}$ in s$^{-1}$),
 oscillator strengths (f$_{ij}$, dimensionless), and line strengths (S, in atomic units) for electric dipole (E1), and 
A$_{ji}$ for electric quadrupole (E2), magnetic dipole (M1), and magnetic quadrupole (M2) transitions of Yb~LXII.
  The ratio R(E1) of velocity and length forms of A-values for E1 transitions is listed in the last column.}
\begin{tabular}{@{}p{1in}p{6in}@{}}
$i$ and $j$         & The lower ($i$) and upper ($j$) levels of a transition as defined in Table 16.\\
$\lambda_{ij}$      & Transition wavelength (in ${\rm \AA}$) \\
A$^{E1}_{ji}$       & Radiative transition probability (in s$^{-1}$) for the E1 transitions \\
f$^{E1}_{ij}$       & Absorption oscillator strength (dimensionless) for the E1 transitions \\
S$^{E1}$            & Line strength in atomic unit (a.u.), 1 a.u. = 6.460$\times$10$^{-36}$ cm$^2$ esu$^2$ for the E1 transitions \\
A$^{E2}_{ji}$       & Radiative transition probability (in s$^{-1}$) for the E2 transitions \\
A$^{M1}_{ji}$       & Radiative transition probability (in s$^{-1}$) for the M1 transitions \\
A$^{M2}_{ji}$       & Radiative transition probability (in s$^{-1}$) for the M2 transitions \\
R(E1)                     & Ratio of velocity and length forms of A- (or f- and S-) values for the E1 transitions \\
$a{\pm}b$ &  $\equiv a\times{10^{{\pm}b}}$ \\
\end{tabular}
\label{ExplTable35}

\bigskip
\section*{Table 36.\label{tbl36te}  Transition wavelengths ($\lambda_{ij}$ in $\rm \AA$), radiative rates (A$_{ji}$ in s$^{-1}$),
 oscillator strengths (f$_{ij}$, dimensionless), and line strengths (S, in atomic units) for electric dipole (E1), and 
A$_{ji}$ for electric quadrupole (E2), magnetic dipole (M1), and magnetic quadrupole (M2) transitions of Lu~LXIII.
  The ratio R(E1) of velocity and length forms of A-values for E1 transitions is listed in the last column.}
\begin{tabular}{@{}p{1in}p{6in}@{}}
$i$ and $j$         & The lower ($i$) and upper ($j$) levels of a transition as defined in Table 17.\\
$\lambda_{ij}$      & Transition wavelength (in ${\rm \AA}$) \\
A$^{E1}_{ji}$       & Radiative transition probability (in s$^{-1}$) for the E1 transitions \\
f$^{E1}_{ij}$       & Absorption oscillator strength (dimensionless) for the E1 transitions \\
S$^{E1}$            & Line strength in atomic unit (a.u.), 1 a.u. = 6.460$\times$10$^{-36}$ cm$^2$ esu$^2$ for the E1 transitions \\
A$^{E2}_{ji}$       & Radiative transition probability (in s$^{-1}$) for the E2 transitions \\
A$^{M1}_{ji}$       & Radiative transition probability (in s$^{-1}$) for the M1 transitions \\
A$^{M2}_{ji}$       & Radiative transition probability (in s$^{-1}$) for the M2 transitions \\
R(E1)                     & Ratio of velocity and length forms of A- (or f- and S-) values for the E1 transitions \\
$a{\pm}b$ &  $\equiv a\times{10^{{\pm}b}}$ \\
\end{tabular}
\label{ExplTable36}

\bigskip
\section*{Table 37.\label{tbl37te}  Transition wavelengths ($\lambda_{ij}$ in $\rm \AA$), radiative rates (A$_{ji}$ in s$^{-1}$),
 oscillator strengths (f$_{ij}$, dimensionless), and line strengths (S, in atomic units) for electric dipole (E1), and 
A$_{ji}$ for electric quadrupole (E2), magnetic dipole (M1), and magnetic quadrupole (M2) transitions of Hf~LXIV.
  The ratio R(E1) of velocity and length forms of A-values for E1 transitions is listed in the last column.}
\begin{tabular}{@{}p{1in}p{6in}@{}}
$i$ and $j$         & The lower ($i$) and upper ($j$) levels of a transition as defined in Table 18.\\
$\lambda_{ij}$      & Transition wavelength (in ${\rm \AA}$) \\
A$^{E1}_{ji}$       & Radiative transition probability (in s$^{-1}$) for the E1 transitions \\
f$^{E1}_{ij}$       & Absorption oscillator strength (dimensionless) for the E1 transitions \\
S$^{E1}$            & Line strength in atomic unit (a.u.), 1 a.u. = 6.460$\times$10$^{-36}$ cm$^2$ esu$^2$ for the E1 transitions \\
A$^{E2}_{ji}$       & Radiative transition probability (in s$^{-1}$) for the E2 transitions \\
A$^{M1}_{ji}$       & Radiative transition probability (in s$^{-1}$) for the M1 transitions \\
A$^{M2}_{ji}$       & Radiative transition probability (in s$^{-1}$) for the M2 transitions \\
R(E1)                     & Ratio of velocity and length forms of A- (or f- and S-) values for the E1 transitions \\
$a{\pm}b$ &  $\equiv a\times{10^{{\pm}b}}$ \\
\end{tabular}
\label{ExplTable37}

\bigskip
\section*{Table 38.\label{tbl38te}  Transition wavelengths ($\lambda_{ij}$ in $\rm \AA$), radiative rates (A$_{ji}$ in s$^{-1}$),
 oscillator strengths (f$_{ij}$, dimensionless), and line strengths (S, in atomic units) for electric dipole (E1), and 
A$_{ji}$ for electric quadrupole (E2), magnetic dipole (M1), and magnetic quadrupole (M2) transitions of Ta~LXV.
  The ratio R(E1) of velocity and length forms of A-values for E1 transitions is listed in the last column.}
\begin{tabular}{@{}p{1in}p{6in}@{}}
$i$ and $j$         & The lower ($i$) and upper ($j$) levels of a transition as defined in Table 19.\\
$\lambda_{ij}$      & Transition wavelength (in ${\rm \AA}$) \\
A$^{E1}_{ji}$       & Radiative transition probability (in s$^{-1}$) for the E1 transitions \\
f$^{E1}_{ij}$       & Absorption oscillator strength (dimensionless) for the E1 transitions \\
S$^{E1}$            & Line strength in atomic unit (a.u.), 1 a.u. = 6.460$\times$10$^{-36}$ cm$^2$ esu$^2$ for the E1 transitions \\
A$^{E2}_{ji}$       & Radiative transition probability (in s$^{-1}$) for the E2 transitions \\
A$^{M1}_{ji}$       & Radiative transition probability (in s$^{-1}$) for the M1 transitions \\
A$^{M2}_{ji}$       & Radiative transition probability (in s$^{-1}$) for the M2 transitions \\
R(E1)                     & Ratio of velocity and length forms of A- (or f- and S-) values for the E1 transitions \\
$a{\pm}b$ &  $\equiv a\times{10^{{\pm}b}}$ \\
\end{tabular}
\label{ExplTable38}

\end{document}